\documentclass[twocolumn,showpacs,preprintnumbers,amsmath,amssymb]{revtex4}

\usepackage{graphicx}
\newcommand\ket[1]{\mid #1 \rangle}
\newcommand\bra[1]{\langle #1 \mid}
\newcommand\media[1]{\langle #1 \rangle}
\begin{document}
\title{Antiholons in one-dimensional $t$-$J$ models.}
\author{C.\ Lavalle$^{1}$, M.\ Arikawa$^{2}$, S.\ Capponi$^{3}$,
F.F.\ Assaad$^{1,4}$ and
A. Muramatsu$^{1}$}
\affiliation{$^{1}$ Institut f\"ur Theoretische Physik III,
Universit\"at Stuttgart, Pfaffenwaldring 58, D-70550 Stuttgart, Germany.
\\
$^{2}$ Department of Mathematics and Statistics,
University of Melbourne, Victoria 3010, Australia.
\\
$^{3}$ Universit\'e Paul Sabatier, Laboratoire de Physique Quantique, 118
route de Narbonne, 31062 Toulouse, France
\\
$^{4}$ Max-Planck-Institut f\"ur Festk\"orperforschung,
Heisenbergstr.\ 1,
D-70569 Stuttgart, Germany.
}
\begin{abstract}
Using a newly developed hybrid Monte Carlo algorithm for the
nearest-neighbor 
(n.n.) $t$-$J$ model, we show that antiholons identified in the
supersymmetric inverse squared
(IS) $t$-$J$ model are clearly visible in the
electron addition spectrum of the n.n.\ $t$-$J$ model at $J=2t$ and also
for  $J = 0.5t$, a value of experimental relevance.
\end{abstract}
\pacs{71.10.Fd, 71.10.Pm,  71.10.-w}
\maketitle
It is well established that electrons in one-dimensional (1D) metals
generally lead to a Luttinger liquid \cite{haldane81,voit95},
where charge-spin separation (CSS) takes place.
The most direct evidence of CSS was predicted 
for the
spectral function of such systems \cite{meden92,voit93}. 
While experimental
evidence of CSS has accumulated in recent years
\cite{kim96,kobayashi99,claessen02},
an exact theoretical evaluation of the one-particle spectral function
$A(k,\omega)$
could until now only be fully accomplished for the Hubbard model at
$U=\infty$ for 
arbitrary doping on the basis of the Bethe-Ansatz solution
\cite{penc96,favand97}.
However, recent progress was made for spectral properties of the
supersymmetric 
(SuSy) $t$-$J$ model with $1/r^2$ interaction, where beyond the exact
ground state (GS)
\cite{kuramoto91},
the thermodynamics \cite{kuramoto95}, the compact support of $A(k,\omega)$
\cite{ha94},
the single-hole dynamics \cite{kato98}, and the electron addition spectrum
\cite{arikawa01} could be calculated analytically.
In addition to spinons and holons,
the IS SuSy $t$-$J$ model was shown to contain antiholons
with charge $Q=2e$, spin $S=0$, and twice the mass of the holons, 
i.e.\ 
they are not merely charge conjugate to the holons. 

We present in this Letter one-particle spectral functions for the 1D n.n.\
$t$-$J$ model with finite doping, obtained by quantum Monte Carlo (QMC)
simulations based on a newly developed hybrid algorithm.
We show that the electron addition part of $A(k,\omega)$ presents a clear
structure following the antiholon dispersion found in
$A(k,\omega)$ of the IS SuSy $t$-$J$ model for the same doping.
Our results show moreover, that also away from that point,
a corresponding feature is present, strongly indicating that
antiholons originally identified in the IS SuSy $t$-$J$
model are generically present in the n.n.\ model.

\begin{figure}[ht]
\begin{picture}(0,0)(125,10)%
\includegraphics[width=.4\textwidth]{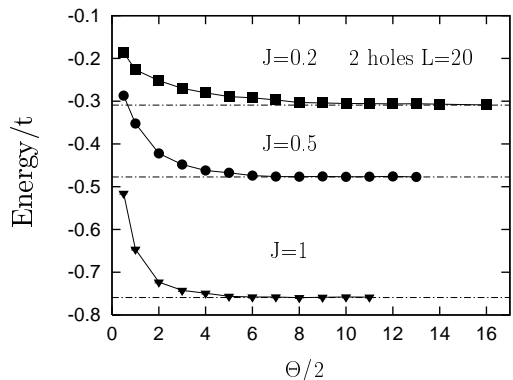}
\end{picture}
\caption[]{Ground-state energies $vs$. exact diagonalization results
for $L=20$ with two holes as a function of the projection parameter
$\Theta$.}
\label{energies}
\end{figure}

The $t$-$J$ model reads:
\begin{eqnarray} 
H_{t-J}&=&
-\sum\limits_{i<j,\sigma} t_{ij} (\tilde c^{\dagger}_{i,\sigma}
\tilde c^{}_{j,\sigma}+{\rm h.c.}) \nonumber \\
& &
 + \sum\limits_{i<j} J_{ij}
\left( {\vec S}_i\cdot {\vec S}_j -\frac 1 4 \tilde n_i \tilde n_j \right).
\label{tJ}
\end{eqnarray}
Here, $\tilde c^{\dagger}_{i,\sigma}$ are projected fermion operators
$\tilde c^{\dagger}_{i,\sigma}=(1-n_{i,-\sigma})c^{\dagger}_{i,\sigma}$,
${\vec S}_i=(1/2)\sum_{\alpha,\beta}c^{\dagger}_{i,\alpha}
{\vec \sigma}_{\alpha,\beta}c^{}_{i,\beta}$,
$\tilde n_i=\sum_{\alpha} \tilde c^{\dagger}_{i,\alpha}\tilde
c^{}_{i,\alpha}$
and $c^{\dagger}_{i,\alpha}$ and $c^{}_{i,\alpha}$ are canonical creation
and annihilation fermionic operators respectively with
$n_{i,\sigma}= c^{\dagger}_{i,\sigma} c^{}_{i,\sigma}$.
We consider two type of interactions: (i) n.n.\ type: $t_{ij}=t
\delta_{j,i+1}$, 
$J_{ij}= J \delta_{j,i+1}$
and (ii) IS SuSy type: $t_{ij}=J_{ij}/2=t (\pi/L)^2/\sin^2(\pi(i-j)/L)$.
$L$ is the length of the system and the lattice constant is unity.
For the n.n.\ model, it is convenient for the development of the algorithm
to 
perform first
a canonical transformation \cite{khaliullin90,antimo95}
\begin{equation}
c^\dagger_{i\uparrow} = \gamma^{+}_{i} f_i - \gamma^{-}_{i} f_i^\dagger \, ,
\; \; \;
c^\dagger_{i\downarrow} = \sigma^{-}_{i} (f_i + f_i^\dagger) \, ,
\label{canonical}
\end{equation}
where $f^{\dagger}_i, f_i$ are canonical operators for spinless fermions,
$\sigma^{\pm}=1/2(\sigma^x \pm i \sigma^y)$, and
$\gamma^{\pm}= 1/2 (1 \pm \sigma^z)$, with $\sigma^{\alpha},\ \alpha=x, y,
z$
Pauli matrices.
The Hamiltonian (\ref{tJ}) becomes:
\begin{equation}
{\cal H} = +t \sum\limits_{<i,j>} P_{ij}f^{\dagger}_if^{}_j +
\frac J 2 \sum\limits_{<i,j>}\Delta_{ij}(P_{ij}-1),
\label{newtJ}
\end{equation}
where $<i,j>$ means n.n.
$P_{ij} = 1/2(1 + {\vec \sigma}_i \cdot {\vec \sigma}_j)$
and $\Delta_{ij} = 1 - f^{\dagger}_i f_i - f^{\dagger}_j f_j$.
The constraint against double occupancy becomes
$\sum_i (1 - \sigma^z_i) f^{\dagger}_i f_i = 0$, and commutes with the
Hamiltonian.
We consider now the following definition of an expectation value:
\begin{equation}
\media{\hat {\cal O}}\ = \lim_{\Theta\rightarrow\infty}\
\frac{\sum_n
\bra{\Psi_n} {\cal P} \
{\rm e}^{-\frac{\Theta}{2} {\cal H}}\ \hat {\cal O} \
{\rm e}^{-\frac{\Theta}{2} {\cal H}}\ {\cal P}
\ket{\Psi_n}}{\sum_n
\bra{\Psi_n} {\cal P} \ {\rm e}^{-\Theta {\cal H}}\
{\cal P} \ket{\Psi_n}} \; ,
\label{media}
\end{equation}
where $\ket{\Psi_n} = \ket{s_n} \ \otimes \ket{\Psi_T}$, with
$\left\{ \ket{s_n} \right\}$ a complete set of spin states and
$\ket{\Psi_T}$ a trial wavefunction for the spinless fermions.
${\cal P}$ is a projector
ensuring the constraint against double occupancy.
Taking the limit $\Theta\rightarrow\infty$ leads each state
${\cal P} \ket{\Psi_T} \ \otimes \ket{s_n}$ to converge to the GS
as long as the GS has a finite overlap with it.
The multiplicity is corrected by the normalization factor.
Introducing after slicing in imaginary time
(typically time slices $\Delta \tau = 0.1/t $ are used)
a complete set of spin states,
and checkerboarding, the spin states are represented by world-lines,
and for each configuration of them, 
fermions are evolved exactly since the Hamiltonian (\ref{newtJ}) is bilinear 
in fermions. 
It can be easily shown that the total weight for a given configuration of
the world-lines is given by $W_H D_f$, where $W_H$ is the weight of an
antiferromagnetic Heisenberg model (AFHM), whereas $D_f$ is a fermionic
determinant \cite{muramatsu99,lavalle02}.
The updating of spin world-lines is performed using the loop-algorithm
\cite{evertz93}, with the same complexity as for an AFHM, in contrast to
a recently proposed pure loop-algorithm \cite{ammon98}. In fact, the 
autocorrelation time ($\tau \sim 2$ for $L=30$ and $J/t=2$) for the 
internal energy is very similar to the one for the AFHM.   
Due to the mixed character we denominate the whole hybrid-loop algorithm.
Figure \ref{energies} shows a comparison of GS energies from QMC and 
exact diagonalization for various values of $J$ at a density $n=0.9$.
The correct value is reached for values of the projection parameter
$\Theta  \sim 10/t - 20/t$,
demonstrating that the algorithm leads to the correct GS with high
accuracy (statistical errors are smaller than the size of the symbols).
Dynamical data are obtained from the imaginary time Green's function
and analytically continued  using the maximum entropy method
\cite{jarrell96}.
Further details will be presented elsewhere \cite{lavalle02}.
\vspace*{4.5cm}

\begin{figure}[hb]
\begin{picture}(0,0)(130,5)%
\includegraphics[width=.5\textwidth]{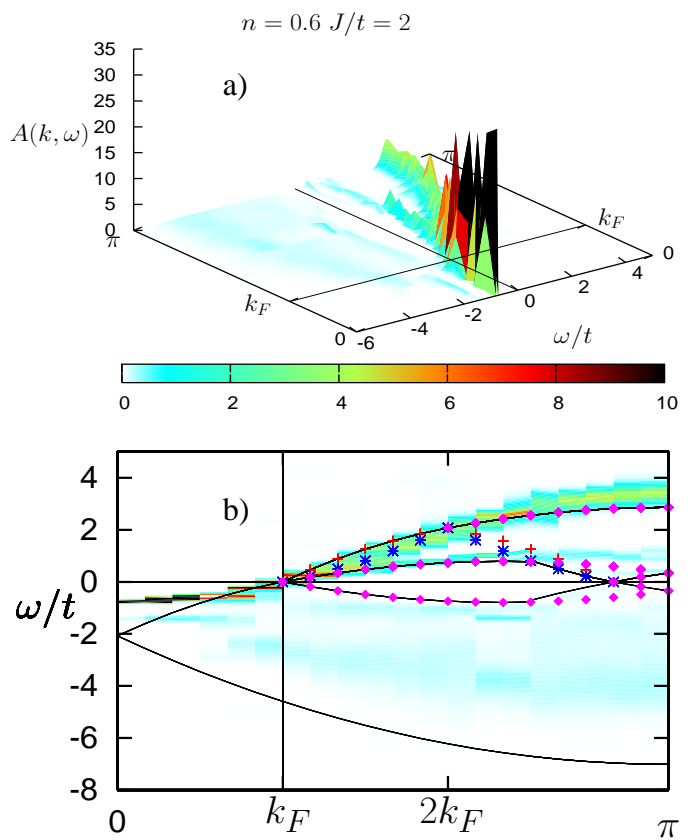}
\end{picture}
\caption[]{a) $A(k,\omega)$ for $J=2t$ at a density $n=0.6$.
b) Projection of intensities on the $(\omega,k)$ plane.
Solid lines: compact support of the IS SuSy $t$-$J$ model.
Red crosses: spinons, blue asterisks: holons, magenta diamonds: antiholons.
See text for the dispersions.}
\label{surfJ2}
\end{figure}

We consider $A(k,\omega)$ both for electron-removal
(ER) and electron-addition (EA) processes.
Figure \ref{surfJ2} a) shows $A(k,\omega)$ obtained from the
QMC simulations for $J=2t$ ($L=40$, $\Theta=24/t$), at a density
$n=0.6$. The Fermi energy is taken as the zero of the energy scale.
A splitting of the spectral weight into two branches can be readily seen
on the EA side, in contradiction with what is expected for a single band.
Figure \ref{surfJ2} b) shows the projection of $A(k,\omega)$ on the
($\omega,k$) plane, revealing the dispersion of the main features in the
spectrum, together with the compact support for the IS
SuSy $t$-$J$ model at the same density. Furthermore, the dispersions of
spinon 
($s$), holon ($h$), and antiholon ($\bar h$) branches that determine the
compact support 
for EA processes in the IS SuSy $t$-$J$ model are also shown.
The dispersions for right ($R$) and left ($L$) going spinons and holons
are given by 
$\epsilon_{\rm sR(L)} (q)/t  =   q (\pm v^0_{\rm s}-q  )$, and
$\epsilon_{\rm hR(L)} (q)/t  =   q ( q \pm v^0_{\rm c} ) $,
respectively, where $v^0_{\rm c} = \pi(1-{n})$ and $v^0_{\rm s}=\pi$.
The antiholon dispersion is
$\epsilon_{\rm \bar{h} } (q)/t  = q \left( 2 v^0_{\rm c} -q \right)/2$.
The accessible range of momenta is for $\epsilon_{s R (L)}$ and
$\epsilon_{h R (L)}$,
$0 \leq q \leq k_F$ ($-k_F \leq q \leq 0$),
and for $\epsilon_{\bar h}$,
$ 0 \leq q \leq 2 \pi - 4 k_F$ \cite{ha94,kuramoto95,arikawa01}.
The compact support is obtained by assuming that the energy and momenta of the
particle (EA) or hole (ER) are given by the addition of energy and momenta of 
$s$, $h$, and $\bar h$ with the dispersions above \cite{ha94}.
In the ER part of the spectrum, only the corresponding part of the compact
support and the dispersion of an antiholon branch along the support is shown
for clarity, since in contrast to the EA processes, where only one spinon,
one
holon, and one antiholon are present, in  the ER part three spinon and holon
contributions \cite{ha94} are possible.  Therefore, a large number of
features would appear, whose intensity is at the moment unknown, and hence,
their importance is difficult to assess.
A sharp feature is visible on the ER side that escapes from the compact
support of the IS SuSy model. It is due to a holon branch, and as already
discussed in the limit of a single hole \cite{brunner00a}, the actual
dispersion
of the holon is needed, in order to describe this feature correctly. Also a
deviation from the IS SuSy compact support is observed on the
EA side at high energies, where the differences in the models is expected to
become noticeable. 
\vspace*{12cm}
There are however, several features that are well described by the 
excitations of the IS model. The strongest feature on the EA 
side is followed closely by the spinon and holon branches between $k_F$ and
$2 k_F$, and for $k>2 k_F$ by a spinon at $k_F$ together with a dispersing
antiholon. The analytic results for EA processes in the IS SuSy model
\cite{arikawa01} show that the largest portion of spectral weight is along
this line. 
More striking is a second, weaker, but clearly visible branch that follows
very closely the dispersion of an antiholon between $k_F$ and $2 \pi - 3 k_F$.
The analytic results of $A(k,\omega)$ for the IS SuSy model 
\cite{arikawa01,arikawa02}
predict a stepwise discontinuity at this edge and, in fact, the explicit 
evaluation of the weight shows for the present range of doping a higher 
value than in the interior of the support. 
Also the upper edge of the compact
support on the ER part, that in the IS model corresponds to an antiholon, is
well reproduced, with spectral weight down to very low energies around $3
k_F$,
as predicted by the IS SuSy model.
Therefore, at the SuSy point,
the clearest signal of CSS in $A(k,\omega)$ for the n.n.\ $t$-$J$ model
are present in the EA part of the spectrum and through the comparison
with the IS $t$-$J$ model, it is clear that a sizeable part of the spectral
weight goes to the antiholon excitation.
\vspace*{12cm}
\begin{figure}[hb]
\begin{picture}(0,0)(130,5)%
\includegraphics[width=.5\textwidth]{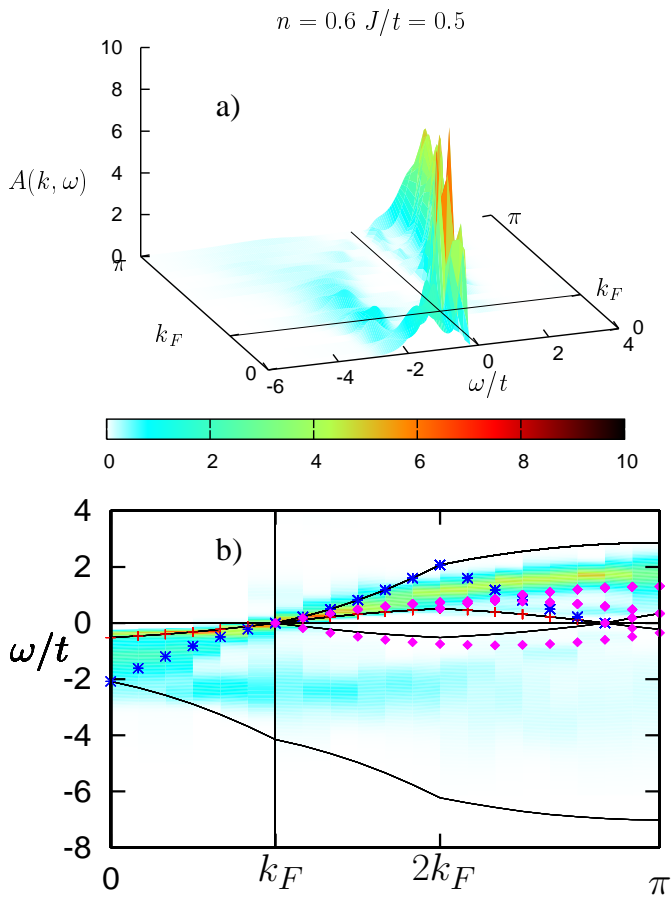}
\end{picture}
\caption[]{a) $A(k,\omega)$ for $J=0.5t$ at $n=0.6$. b) Projection of
intensities on the $(\omega,k)$ plane. Symbols coded as in Fig.\
\ref{surfJ2}}
\label{surfJ0.5}
\end{figure} 
Further analytical results con\-cer\-ning the detailed structure of 
$A(k,\omega)$ on the EA part and for the one-holon and one-spinon 
contributions on the ER part will be published elsewhere \cite{arikawa02}.  

Since the exact solution of the IS model is restricted to the SuSy point, it
is of much interest to see whether the features discussed above correspond
to a
generic behavior of the n.n.\ $t$-$J$ model or whether it
is better described, e.g.\ by the solution of the n.n.\ model at $J=0$
\cite{penc96}.
Figure \ref{surfJ0.5} a) shows $A(k,\omega)$ for
$n=0.6$ and $J=0.5t$ ($L=40$, $\Theta = 56/t$), i.e.\ very far away from
the SuSy point and at a value of $J/t$
of experimental relevance for cuprate compounds.
A perspective was chosen, so that it is
already visible that as in the SuSy case, a structure splits off the main
feature for $k$ between $2 k_F$ and $\pi$. Figure \ref{surfJ0.5} b)
shows the projection of
$A(k,\omega)$ on the ($\omega,k$) plane.
As a model for free spinons, holons, and antiholons, we use the same
dispersions as for the IS SuSy model, but with
$\epsilon_{\rm sR(L)} (q)  =  (J/2)  q (\pm v^0_{\rm s}-q  )$, i.e.\
assuming
that away from the SuSy point, only the energy scale of spinons is changed.
The corresponding compact support, spinon, holon, and antiholon dispersions
are encoded as in Fig.\ \ref{surfJ2}. In the present case, the compact
support 
encloses rather well all the spectral weight. Moreover, on the ER part, the
strongest feature is very accurately followed by a spinon, whereas a second
structure is also closely followed by a holon. A more detailed view of these
structures is given below in Fig.\ \ref{Spinon-holon}.
They correspond to the generally expected signal in photoemission
for CSS, that were also found
in previous numerical studies of the Hubbard model \cite{zacher98}.
However, as shown in Fig.\ \ref{Spinon-holon}, the present algorithm
seems to lead to results accurate enough, so that after application of
maximum 
entropy, CSS is seen below $E_F$ in a wider range in $k$-space than
previously.
On the EA side, the feature with largest intensity is followed close to
$k_F$
by a holon, a spinon, and an antiholon. However, further away from $k_F$,
the dispersion of the maximum is, up to $k \sim 2 k_F$,
closer to an antiholon going from $k_F$ to $2 \pi - 3 k_F$ and beyond $2
k_F$
by a curve corresponding to a spinon at $k_F$ and a dispersing antiholon.
Moreover,
a second maximum develops beyond $2 k_F$ that follows the antiholon
dispersing from $k_F$ to $2 \pi - 3 k_F$,
in a similar way as for $J=2t$ but with a smaller gap between both curves.
In particular, the results from the simulations show appreciable weight
between $3 k_F$ and the zone boundary, where only antiholons are present.
  
A closer look to both features signaling CSS is given 
in Fig.\ \ref{Spinon-holon}.
Figure \ref{Spinon-holon} a) shows $A(k,\omega)$ on the ER side
and the location of
the excitation energies for one spinon and one holon.
Whereas the spinon dispersion follows the QMC data very closely, a deviation
is seen for the holon for the farthest points from $k_F$, as can be
expected, 
since at higher energies, details of the dispersion matter in general. Yet,
the 
agreement is good enough to enable an identification of the excitation
content
of the spectrum.
The details of the splitted maxima for $ 2 k_F \leq k \leq \pi$
on the EA side are shown in
Fig.\ \ref{Spinon-holon} b), where both an antiholon dispersing from $k_F$
to
$2 \pi - 3 k_F$ (closer to $\omega=0$) and an antiholon dispersing from $2
k_F$
to $2 \pi - 2 k_F$ on top of a spinon at $k_F$ are shown. Whereas the latter
follows the larger maximum, the former can be associated with the second
maximum. 
As at the SuSy point, there seems to be almost no weight associated with
the left propagating spinon and holon that give rise to the contributions
between $2 k_F$ and $3 k_F$. This is consistent with the analytic results
obtained for the IS SuSy model \cite{arikawa01}.
Results for other values of doping ($0.6 \leq n \leq 0.9$) and $J$
($0.5 \leq J/t \leq 3$) not presented here, show the same qualitative
behavior,
in particular the presence of a branch on the EA side below the main
dispersing structure, that is closely followed by an antiholon branch
under the assumption of free spinons, holons, and antiholons.
\vspace*{6cm}
\begin{figure}[ht]
\begin{picture}(0,0)(130,10)%
\includegraphics[width=.5\textwidth]{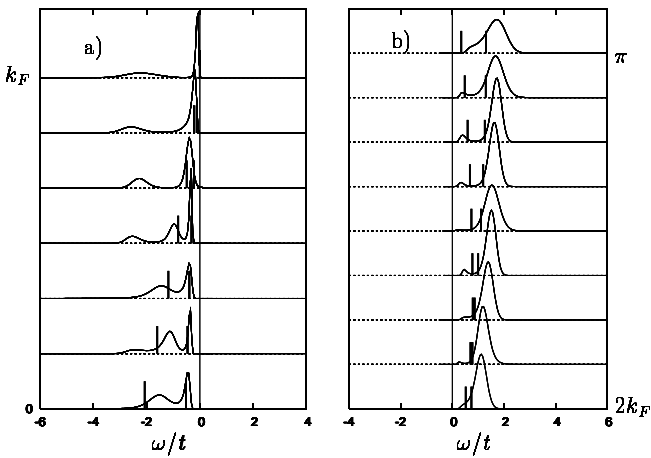}
\end{picture}
\caption[]{Detailed view of $A(k,\omega)$ at $J=0.5t$ and
$n=0.6$.
a) ER side for $0 \leq k \leq k_F$ at $J=0.5t$ and
$n=0.6$. Vertical bars denote the positions for free spinon (closer to
$\omega=0$) and holon excitations. b) EA side for
$2 k_F \leq k \leq \pi$. Vertical bars denote here antiholon dispersions.}
\label{Spinon-holon}
\end{figure}

In summary, one-particle spectra for electron-removal and addition were
obtained using a new algorithm that delivers accurate dynamical data for the
nearest-neighbor $t$-$J$ model.
A comparison with the compact support and excitation content of the $1/r^2$
$t$-$J$ model at the supersymmetric point $J=2t$ shows that a new
manifestation of charge-spin separation in the n.n.\ model
can be observed in the EA part of the spectrum, where
in addition to spinons and holons, 
a branch following the antiholon dispersion is clearly visible. 
The same
feature is still visible at $J=0.5t$, where assuming the same
dispersions for the holon, and antiholon, as in the IS model
but changing the scale of energy to $J$ for the spinon,
a fairly good description of the spectrum can be given.
Instead, serious deviations result by omiting the antiholon or
setting its mass equal to that of the holon (i.e.\ assuming 
that it is the charge conjugated counterpart of the holon) \cite{lavalle02}.
The results above strongly indicate, that
antiholons, that are not charge conjugate of holons,
are generic excitations in the nearest neighbor $t$-$J$ model.

We are grateful to M. Brunner for important contributions at early stages 
of this project. 
We wish to thank HLR-Stuttgart (Project DynMet) and HLRZ-J\"ulich for
allocation of computer time and SFB 382 for financial support.
M.A. acknowledges support by the Visitor Program of the MPI-PKS
and Australian Research Council. A.M.\ is grateful to the ITP, Santa
Barbara, for its kind hospitality, and
support in part under NSF Grant No PHY99-07949.

\end{document}